\documentclass{article} 
\usepackage[final]{colm2026_conference}

\usepackage{microtype}
\usepackage{graphicx}
\usepackage{booktabs}
\usepackage{tabularx}
\usepackage{multirow}
\usepackage[table]{xcolor}
\usepackage{adjustbox}
\usepackage{float}
\usepackage{rotating}            
\usepackage{amsmath}
\usepackage{amssymb}
\usepackage{pifont}              
\usepackage[ruled,vlined]{algorithm2e}
\usepackage{tikz}
\usepackage{url}
\usepackage{hyperref}            
\usepackage{cleveref}

\newcommand{\cmark}{\ding{51}}
\newcommand{\xmark}{\ding{55}}
\providecommand{\ninept}{\fontsize{9pt}{11pt}\selectfont}
\newcolumntype{Y}{>{\centering\arraybackslash}X} 

\title{MetaSICL: Globalizing Auditory LLMs for Underserved Speakers and Languages via Meta Speech In-Context Learning}

\author{Haolong Zheng\thanks{Equal contribution.} \\
University of Illinois Urbana-Champaign \\
\texttt{haolong2@illinois.edu}
\And
Siyin Wang\footnotemark[1] \\
Tsinghua University \\
\texttt{wangsiyi23@mails.tsinghua.edu.cn}
\And
Zengrui Jin \\
Tsinghua University \\
\texttt{
zengrui.jin0@gmail.com
}
\And
Mark Hasegawa-Johnson \\
University of Illinois Urbana-Champaign \\
\texttt{jhasegaw@illinois.edu}
}

\begin{document}

\maketitle

\begin{abstract}
Generative AI systems for speech and audio are increasingly expected to serve users across languages, cultures, and communities, yet current auditory Large Language Models (LLMs) are still largely trained and evaluated on high-resource data. Globalizing such systems therefore requires overcoming low-resource settings, where the target speakers, languages, or tasks are poorly represented in training data. In these regimes, collecting enough labeled in-domain data is often impractical, and the small corpora that can be collected may still under-represent the true test distribution, making direct fine-tuning brittle under domain shift. In-Context Learning (ICL) offers a promising alternative: instead of updating model parameters for every underserved community, an auditory LLM can adapt at inference time by conditioning on only a few local demonstrations. However, vanilla speech ICL remains limited because most auditory LLMs are not explicitly trained to use such demonstrations effectively. We address this gap with \textbf{Meta Speech In-Context Learning (MetaSICL)}, a post-training recipe that strengthens an auditory LLM's in-context adaptation ability using only abundant high-resource speech data. Although MetaSICL never trains on the target low-resource domains, it improves performance across two model backbones on children's ASR, audio understanding/reasoning, and speech translation and ASR in directions and languages not included in its post-training data. We further study the setting where a small amount of in-domain data is available, using low-resource language ASR as a case study, since speech recognition for underserved languages is central to globalizing generative AI. In this setting, using MetaSICL as a warmup for in-domain reinforcement learning yields the strongest results, outperforming direct fine-tuning across five typologically diverse languages. Overall, MetaSICL offers a practical route toward globalizing auditory LLMs by building inference-time adaptation into the model.
\end{abstract}

\begin{figure*}[ht]
    \centering
    \includegraphics[width=1\linewidth]{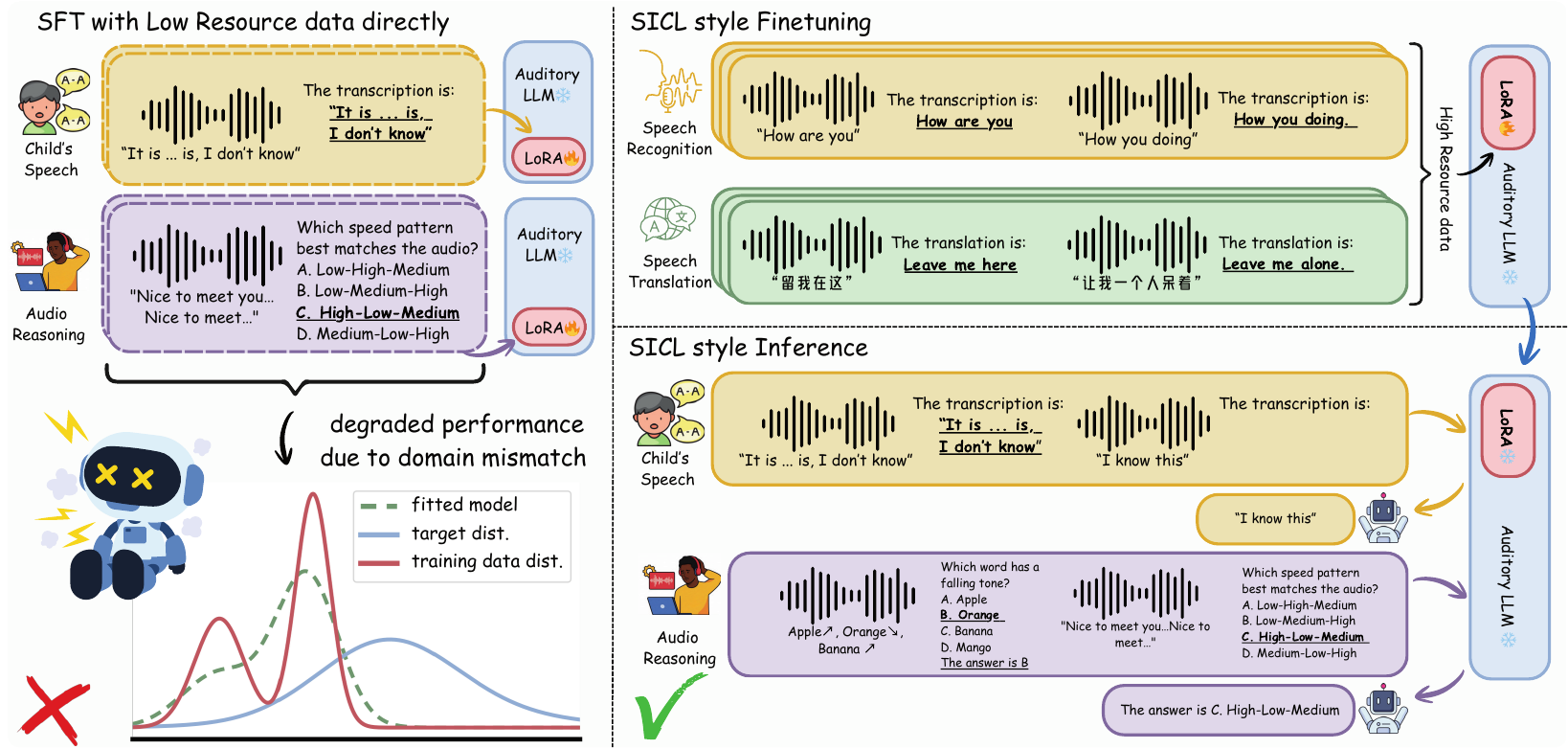}
    \caption{Motivation and overview of MetaSICL for low-resource audio tasks. Left: Direct supervised fine-tuning (SFT) on limited in-domain data often degrades performance under distribution shift due to domain mismatch between scarce training samples and the target test distribution. Right: We instead perform SICL-style fine-tuning on abundant, high-resource speech tasks using an ICL formatted objective, then apply SICL-style inference by providing task demonstrations at test time to adapt the same LoRA-augmented auditory LLM to low-resource domains, improving robustness and downstream performance.}
    \label{fig:overview}
\end{figure*}

\section{Introduction}

Globalizing generative AI requires models that can adapt to low-resource users, languages, and domains. This challenge is especially acute for speech and audio: while high-resource settings such as adult English speech provide abundant labeled data, the vast majority of the world's languages, dialects, speaking styles, age groups, and culturally specific interaction patterns remain sparsely represented. As a result, globally useful auditory LLMs cannot assume that every target community will have enough in-domain supervision for conventional fine-tuning. Progress therefore hinges on \emph{low-resource adaptation}: enabling models trained primarily on high-resource data to adjust to underrepresented speakers, languages, and tasks from only a small amount of local evidence.

However, current auditory LLMs are not yet as robust under low-resource adaptation as global deployment requires. Although they achieve strong results on mainstream speech and audio benchmarks, their performance often drops sharply when applied to underrepresented settings: children's speech differs from adult speech in acoustics, pronunciation, speaking rate, and disfluency; many languages and dialects appear rarely or never during training; many speech-translation directions have little paired speech--text data; and audio understanding/reasoning (AU/AR) in unfamiliar domains may require domain-specific or culturally grounded interpretation. These failures arise in part because current model development is still dominated by high-resource data, with adult English speech serving as a typical example, leaving many target communities poorly covered during training.

Improving these low-resource tasks through standard fine-tuning is also difficult. In-domain data are expensive to collect and annotate, and the small datasets that can be obtained may still be under-representative of the true target distribution. Fine-tuning on such narrow supervision can therefore overfit to dataset-specific artifacts and become brittle, or even harmful, under domain shift. This tension raises a central question for globalizing auditory LLMs:

\begin{center}
\textit{Can low-resource speech tasks benefit from abundant but out-of-domain data?}
\end{center}


Through In-Context Learning (ICL)\citep{brown2020language}, LLMs can be adapted to new tasks by conditioning on a small set of labeled in-domain examples, without requiring gradient updates. This approach has been shown to be effective across a wide range of modalities \citep{huang2023language, kong2024audio, coreteam2025mimoaudio}. Within the speech domain specifically, ICL has demonstrated notable gains on various tasks including automatic speech recognition (ASR) for children’s speech and unseen languages or dialects \citep{wang2024can, wang2024bayesian, zhou2025m2r, zheng2025ticltextembeddingknnspeech, zheng2025ticlcasestudyspeech}, speech translation (ST) \citep{pan2023cosmic, chen2024salm}, and speech emotion recognition \citep{yang2024uniaudio, ihori2025few}. Fundamentally, ICL enables models to explicitly exploit contextual information to guide generation. This enables us to improve performance on low-resource tasks using only a small number of in-domain examples. In this paper, we first demonstrate that vanilla ICL consistently improves performance across diverse speech and audio tasks.

Moreover, can we further improve this ICL adaptation ability with high-resource out-of-domain data? 
MetaICL \citep{min2022metaicl} enhances a textual LLM's ICL ability via few-shot training on various NLP tasks. SMILE \citep{hsu2024smile}, on the other hand, utilizes high-resource-language ASR data to perform ICL-style tuning on Whisper. However, whether a similar paradigm can be applied to auditory LLMs remains unclear.

To bridge this gap, we propose \textbf{Meta Speech In-Context Learning (MetaSICL)}, a meta-training strategy for auditory LLMs. Rather than fine-tuning the model on each target low-resource domain, MetaSICL constructs ICL-style training episodes from high-resource speech tasks and trains the model to predict a query output conditioned on a small set of audio demonstrations. The goal is not merely to improve performance on the high-resource training tasks themselves, but to teach the model \emph{how to use demonstrations} as contextual cues for test-time adaptation.

Using only abundant high-resource speech tasks for this meta-training stage, MetaSICL yields consistent gains on two model backbones across a broad set of low-resource speech and audio settings, including children's ASR, ASR in languages outside the post-training ASR mixture, speech translation directions held out from post-training, and AU/AR. Rather than requiring a dedicated fine-tuning corpus for every target community, MetaSICL builds the capacity to adapt at inference time---from a handful of in-domain examples---into the model itself. We further show that this demonstration-trained initialization is more stable than direct fine-tuning in low-resource scenarios. Finally, we focus on low-resource language ASR as a dedicated case study, since speech recognition for underserved languages is a central task for globalizing generative AI. When a small in-domain corpus \emph{can} be collected, using MetaSICL as a warmup for in-domain reinforcement learning yields the strongest recognizer, outperforming direct fine-tuning across five typologically diverse languages.


Concretely, we make the following contributions:
\begin{itemize}
    \item \textbf{We establish vanilla SICL as a practical low-resource adaptation mechanism for auditory LLMs.} 
    By providing a few retrieved in-domain demonstrations at inference time, vanilla SICL improves performance across diverse speech and audio settings, including children's ASR, audio understanding/reasoning, multilingual ASR, and speech translation.

    \item \textbf{We propose MetaSICL, a meta-training recipe that teaches auditory LLMs how to use demonstrations.} 
    MetaSICL constructs ICL-style training episodes from abundant high-resource speech tasks, rather than from each target low-resource domain, and trains the model to condition its prediction on audio demonstrations. We further analyze how the composition of the meta-training tasks shapes downstream improvements.

    \item \textbf{We study low-resource language ASR as a key case for globalizing generative AI.} 
    When a small in-domain corpus is available, MetaSICL serves as an effective warmup for a data-efficient in-domain reinforcement-learning stage, yielding the strongest ASR performance across five typologically diverse languages and outperforming direct fine-tuning.
\end{itemize}

\section{Methodology}
We propose \textbf{MetaSICL}, a post-training recipe that explicitly teaches an auditory LLM to inference conditioned on a small set of in-context audio demonstrations. Figure~\ref{fig:overview} illustrates the motivation and overall framework. Algorithm~\ref{alg:training-overview} summarizes the training procedure, and Table~\ref{tab:training-data} lists the data used in each stage. We apply MetaSICL to two auditory LLM backbones---Qwen2.5-Omni \citep{Qwen2.5-Omni} and MiMo-Audio \citep{coreteam2025mimoaudio}---updating only lightweight LoRA adapters to avoid overfitting.\footnote{We update LoRA adapters (rank $8$, $\alpha=32$) inserted into the language backbone of each model; all other parameters are frozen.}

\subsection{Training Instance}

MetaSICL uses an episodic training format that mirrors inference-time in-context learning. 
For each task $c$, a \emph{query set} $\mathcal{D}^{(c)}_{\text{query}}$ and a \emph{demonstration pool} $\mathcal{D}^{(c)}_{\text{pool}}$ are maintained. 
A query instance $(x_{\text{query}}, y_{\text{query}}) \sim \mathcal{D}^{(c)}_{\text{query}}$ and retrieve $k$ in-context demonstrations $\{(x_i, y_i)\}_{i=1}^k$ are sampled from $\mathcal{D}^{(c)}_{\text{pool}}$ at each step.
The prompt is then constructed using the concatenation of the $k$ demonstrations followed by the query.
Conditioned on this full context, the model generates the response $y_{\text{query}}$ according to $P\!\left(y_{\text{query}} \mid x_1,y_1,\ldots,x_k,y_k,x_{\text{query}}\right)$.

\subsection{Training Data}
\textbf{MetaSICL} is post-trained on two abundant, high-resource, out-of-domain tasks: Speech Recognition (ASR) and Speech Translation (ST) (Table~\ref{tab:training-data}).
We utilized the English subset of CommonVoice \citep{commonvoice:2020} and the en$\rightarrow$zh, de$\rightarrow$en, zh$\rightarrow$en, pt$\rightarrow$en subsets of CoVoST2 \citep{wang2020covost} for the ASR and ST tasks, respectively.
For both tasks, demonstrations are retrieved from the training split using TICL \citep{zheng2025ticltextembeddingknnspeech}. 
This ASR$+$ST mixture is the default used throughout unless stated otherwise. To study how the training-task mix shapes downstream behaviour, we additionally consider two ablation variants (Section~\ref{sec:ablation-recipe}): an \emph{ASR-only} recipe that drops ST, and a \emph{$+$SQA} recipe that adds Speech Question Answering.
The $+$SQA variant uses the MMSU dataset \citep{wang2025mmsu}; since MMSU lacks official splits and is relatively small, only the query instance is excluded from the demonstration pool in a ``leave-one-out'' manner.

\begin{table*}[tbp]
\centering
\ninept
\renewcommand{\arraystretch}{1.12}
\setlength{\tabcolsep}{3pt}

\caption{Main results on both backbones. For each model we report zero-shot inference, few-shot inference with retrieved demonstrations (\emph{Vanilla SICL}), and few-shot inference after our post-training (\emph{MetaSICL}, trained only on high-resource ASR$+$ST). Multilingual ASR and ST use CommonVoice and CoVoST2 subsets not included in our post-training data. AU/AR (MMAU, MMAR) accuracy is in percent. Best per column within each backbone in \textbf{bold}; per-task MMAU/MMAR breakdowns are in Appendix~\ref{appendix:AU-AR}.}
\label{tab:results_merged}

\begin{tabularx}{\linewidth}{@{}>{\raggedright\arraybackslash}p{2.4cm} c *{9}{Y}@{}}
\toprule
& 
& \multicolumn{2}{c}{Child's ASR}
& \multicolumn{2}{c}{AU/AR}
& \multicolumn{3}{c}{Multilingual ASR}
& \multicolumn{2}{c}{ST} \\
\cmidrule(l{2pt}r{2pt}){3-4}
\cmidrule(l{2pt}r{2pt}){5-6}
\cmidrule(l{2pt}r{2pt}){7-9}
\cmidrule(l{2pt}r{2pt}){10-11}
& \shortstack{Few-\\shot} 
& \multicolumn{2}{c}{$\downarrow$WER}
& \multicolumn{2}{c}{$\uparrow$Acc.}
& \multicolumn{3}{c}{$\downarrow$WER}
& \multicolumn{2}{c}{$\uparrow$BLEU} \\
\midrule
Tasks &  & MyST & RSR & MMAU & MMAR & de & zh & fr & en$\rightarrow$ja & ja$\rightarrow$en \\
\midrule

\rowcolor{white}
\textit{MiMo-Audio}   & \xmark  & 14.25 & 31.39 & 66.90 & 54.70 & 69.74 & 14.43 & 90.68 & 5.25 & 4.56 \\
\rowcolor{gray!15}
+Vanilla SICL         & \cmark  & 11.55 & \textbf{16.84} & 72.60 & 58.20 & 37.46 & 11.05 & 45.88 & 26.56 & 13.95 \\
\rowcolor{white}
+MetaSICL      & \cmark  & \textbf{11.51} & 16.89 & \textbf{72.90} & \textbf{61.00} & \textbf{30.49} & \textbf{6.51} & \textbf{39.47} & \textbf{36.92} & \textbf{16.76} \\
\cmidrule[\heavyrulewidth]{1-11}

\rowcolor{white}
\textit{Qwen2.5-Omni} & \xmark  & 23.05 & 35.65 & 65.80 & 49.20 & 8.30 & 7.29 & 11.15 & 33.53 & 16.24 \\
\rowcolor{gray!15}
+Vanilla SICL         & \cmark  & 22.72 & 27.86 & 67.30 & 53.80 & 7.48 & 8.07 & 10.39 & 33.65 & 17.47 \\
\rowcolor{white}
+MetaSICL       & \cmark  & \textbf{17.03} & \textbf{21.95} & \textbf{71.10} & \textbf{54.40} & \textbf{7.07} & \textbf{7.18} & \textbf{9.26} & \textbf{35.72} & \textbf{18.15} \\
\bottomrule
\end{tabularx}
\end{table*}

\subsection{Evaluation Data \& Metrics}
Our evaluation deliberately targets the conditions where auditory LLMs are weakest: rather than reporting a single aggregate score, we organize it around low-resource tasks and languages and report breakdowns (Appendix~\ref{appendix:AU-AR}) that expose differences a single headline number would hide.
Child's ASR and Audio Understanding/Reasoning are used to evaluate ICL capability under low-resource scenarios, where the target users (children, and listeners reasoning over culturally diverse audio) are poorly covered by standard adult, English-centric training data.
Non-overlapping tasks spanning multilingual ASR and speech translation---in languages and translation directions not included in our MetaSICL post-training data---are further incorporated to validate that the adaptation ability generalizes across languages rather than merely improving English recognition.
\paragraph{Child's ASR.}
Performance of child ASR is evaluated on two corpora with distinct distributions: My Science Tutor (MyST) \citep{pradhan2024my} and Redmond Sentence Recall (RSR) \citep{ai4exceptionaled_rsr_hf}. 
MyST contains conversational speech from Grade 3–5 students, whereas RSR comprises scripted speech from children aged 5–9. 
Following \citet{zheng2025interspeech}, the utterance-level word error rate (WER) is computed, capped at 1, and then averaged across utterances to mitigate the impact of severe hallucinations on the aggregate metric.

\paragraph{Audio Understanding/Reasoning.}


Performance of general audio understanding and reasoning is evaluated on MMAU and MMAR, which encompass speech, ambient sound, and music comprehension tasks. 
MMAU includes a diverse set of tasks covering 27 skills, with a focus on perception and domain-specific reasoning. 
MMAR comprises 16 subcategory tasks spanning speaker, environment, and content reasoning; audio quality and difference assessment; music and aesthetics; anomaly, spatial, and temporal analysis; and general reasoning. 
For both datasets, accuracy is reported on the public test split using the official evaluation scripts.

\paragraph{Multilingual ASR \& Speech Translation.}

To verify that the improvements generalize beyond the training tasks, multilingual ASR and ST are additionally evaluated on CommonVoice \citep{commonvoice:2020} and CoVoST2 \citep{wang2020covost}, respectively. 
Evaluation is conducted on languages and translation pairs not included in our post-training data (de, fr, zh, en$\rightarrow$ja, and ja$\rightarrow$en).
Word error rate (WER) is reported for alphabetic languages, character error rate (CER) for Chinese, and BLEU with up to 4-gram precision for ST performance.

\begin{table*}[tbp]
\centering
\ninept
\renewcommand{\arraystretch}{1.12}
\setlength{\tabcolsep}{3pt}
\caption{\textbf{Ablation: training-task composition.} From the default ASR$+$ST recipe we drop ST (\emph{ASR-only}) or add SQA (\emph{$+$SQA}); all variants are evaluated few-shot. AU/AR (MMAU, MMAR) accuracy is reported in percent. Best per column within each backbone in \textbf{bold}.}
\label{tab:ablation-recipe}
\begin{tabularx}{\linewidth}{@{}>{\raggedright\arraybackslash}p{2.7cm} c *{9}{Y}@{}}
\toprule
&
& \multicolumn{2}{c}{Child's ASR}
& \multicolumn{2}{c}{AU/AR}
& \multicolumn{3}{c}{Multilingual ASR}
& \multicolumn{2}{c}{ST} \\
\cmidrule(l{2pt}r{2pt}){3-4}
\cmidrule(l{2pt}r{2pt}){5-6}
\cmidrule(l{2pt}r{2pt}){7-9}
\cmidrule(l{2pt}r{2pt}){10-11}
& \shortstack{Few-\\shot}
& \multicolumn{2}{c}{$\downarrow$WER}
& \multicolumn{2}{c}{$\uparrow$Acc.}
& \multicolumn{3}{c}{$\downarrow$WER}
& \multicolumn{2}{c}{$\uparrow$BLEU} \\
\midrule
Recipe &  & MyST & RSR & MMAU & MMAR & de & zh & fr & en$\rightarrow$ja & ja$\rightarrow$en \\
\midrule
\rowcolor{white}
\multicolumn{11}{@{}l}{\textit{MiMo-Audio}} \\
\rowcolor{gray!15}
\quad ASR-only        & \cmark & \textbf{11.49} & \textbf{16.59} & 71.90 & 57.70 & 34.11 & 6.59 & 45.50 & 1.40 & 12.43 \\
\rowcolor{white}
\quad ASR$+$ST (ours) & \cmark & 11.51 & 16.89 & 72.90 & 61.00 & \textbf{30.49} & \textbf{6.51} & \textbf{39.47} & \textbf{36.92} & \textbf{16.76} \\
\rowcolor{gray!15}
\quad $+$SQA          & \cmark & \textbf{11.49} & 16.95 & \textbf{73.40} & \textbf{61.40} & 31.22 & 6.62 & 40.46 & 36.84 & 15.24 \\
\cmidrule[\heavyrulewidth]{1-11}
\rowcolor{white}
\multicolumn{11}{@{}l}{\textit{Qwen2.5-Omni}} \\
\rowcolor{gray!15}
\quad ASR-only        & \cmark & \textbf{14.76} & \textbf{20.96} & 69.60 & 54.20 & \textbf{6.71} & \textbf{6.96} & \textbf{8.80} & 33.56 & 17.39 \\
\rowcolor{white}
\quad ASR$+$ST (ours) & \cmark & 17.03 & 21.95 & 71.10 & 54.40 & 7.07 & 7.18 & 9.26 & \textbf{35.72} & \textbf{18.15} \\
\rowcolor{gray!15}
\quad $+$SQA          & \cmark & 17.42 & 22.16 & \textbf{72.10} & \textbf{54.50} & 7.09 & 7.37 & 9.11 & 34.48 & 17.74 \\
\bottomrule
\end{tabularx}
\end{table*}

\section{Experiments}











\subsection{Vanilla In-Context Learning}

Across both models, adding retrieved demonstrations at inference time (\textit{Vanilla SICL}) improves performance on most benchmarks, indicating that auditory LLMs can leverage in-context examples as a lightweight test-time adaptation signal. Gains are especially consistent on child's ASR and AU/AR, with only a small number of exceptions.

\subsection{Effect of MetaSICL}

With its training mix fixed to ASR$+$ST, \textbf{MetaSICL} delivers the strongest few-shot results across both backbones (Table~\ref{tab:results_merged}). Relative to Vanilla SICL, it sharply improves children's ASR, multilingual ASR in languages outside the post-training mixture, and speech translation in directions held out from post-training, while also lifting general audio understanding and reasoning. Crucially, these gains land on tasks the post-training data never covers: strengthening a model's in-context adaptation does not require large-scale data from the exact downstream task. We analyse \emph{which} training tasks drive \emph{which} downstream gains, and compare against direct fine-tuning, in Section~\ref{sec:ablation}.

\paragraph{A global and cultural view of the gains.}
Beyond aggregate accuracy, the breakdowns in Appendix~\ref{appendix:AU-AR} show that MetaSICL helps precisely where Western-centric, adult-English-centric pipelines fall short, improving MMAR's \emph{Cultural Layer} and \emph{Culture of Speaker} and MMAU's \emph{Socio-cultural Interpretation} categories without any culturally targeted training data. The gains also concentrate on the most distribution-shifted conditions: both backbones improve far more on scripted speech from very young children (RSR, ages 5--9) than on the less-shifted MyST corpus. Demonstration-conditioned adaptation thus disproportionately benefits the users farthest from the training distribution---exactly where collecting large in-domain corpora is least feasible.

\section{Ablation Studies}
\label{sec:ablation}

\subsection{Training-Task Composition}
\label{sec:ablation-recipe}

Holding the model and inference protocol fixed, we vary only the post-training task mix (Table~\ref{tab:ablation-recipe}); the best recipe tracks how closely the training tasks match the downstream task. \textbf{Dropping ST} (ASR-only) gives the strongest ASR but can collapse speech translation: with lexically close ASR/ST instructions, MiMo-Audio reverts to transcribing rather than translating---specializing on one task can silently break another. \textbf{Adding SQA} (MMSU), whose prompt--answer format matches AU/AR, gives the best AU/AR accuracy but slightly degrades ASR and ST. Our default \textbf{ASR$+$ST} is the best overall trade-off (near-best ASR, best ST, strong AU/AR), giving a simple guideline: post-training helps a task most when the training tasks resemble it in supervision and prompt--answer format.

\subsection{Comparison to Direct Fine-tuning}

\begin{table*}[tbp]
\centering
\ninept
\renewcommand{\arraystretch}{1.12}
\setlength{\tabcolsep}{3pt}
\caption{\textbf{Comparison to direct fine-tuning} (Qwen2.5-Omni). Direct supervised fine-tuning on a small in-domain corpus (RSR) or on high-resource same-task data (Common Voice English, CV-en), compared against MetaSICL. AU/AR (MMAU, MMAR) accuracy is in percent. Best child-ASR (MyST/RSR) in \textbf{bold}.}
\label{tab:direct-ft}
\begin{tabularx}{\linewidth}{@{}>{\raggedright\arraybackslash}p{2.5cm} c *{9}{Y}@{}}
\toprule
&
& \multicolumn{2}{c}{Child's ASR}
& \multicolumn{2}{c}{AU/AR}
& \multicolumn{3}{c}{Multilingual ASR}
& \multicolumn{2}{c}{ST} \\
\cmidrule(l{2pt}r{2pt}){3-4}
\cmidrule(l{2pt}r{2pt}){5-6}
\cmidrule(l{2pt}r{2pt}){7-9}
\cmidrule(l{2pt}r{2pt}){10-11}
& \shortstack{Few-\\shot}
& \multicolumn{2}{c}{$\downarrow$WER}
& \multicolumn{2}{c}{$\uparrow$Acc.}
& \multicolumn{3}{c}{$\downarrow$WER}
& \multicolumn{2}{c}{$\uparrow$BLEU} \\
\midrule
Setting &  & MyST & RSR & MMAU & MMAR & de & zh & fr & en$\rightarrow$ja & ja$\rightarrow$en \\
\midrule
\rowcolor{white}
Qwen2.5-Omni            & \xmark & 23.05 & 35.65 & 65.80 & 49.20 & 8.30 & 7.29 & 11.15 & 33.53 & 16.24 \\
\rowcolor{gray!15}
+MetaSICL (ours)        & \cmark & \textbf{17.03} & \textbf{21.95} & 71.10 & 54.40 & 7.07 & 7.18 & 9.26 & 35.72 & 18.15 \\
\rowcolor{white}
Direct FT (CV-en)       & \xmark & 19.83 & 30.61 & 63.90 & 50.50 & 8.34 & 7.44 & 10.49 & 33.25 & 11.90 \\
\rowcolor{gray!15}
\quad +Vanilla SICL     & \cmark & 18.05 & 23.62 & 64.10 & 50.40 & 6.97 & 7.16 & 9.13 & 33.49 & 18.11 \\
\rowcolor{white}
Direct FT (RSR)         & \xmark & 29.47 & 31.09 & 65.30 & 44.90 & 8.43 & 7.84 & 11.42 & 43.72 & 16.27 \\
\bottomrule
\end{tabularx}
\end{table*}

To highlight our approach in low-resource settings, we compare against direct fine-tuning (LoRA, matched to MetaSICL; Table~\ref{tab:direct-ft}). Fine-tuning on the small in-domain RSR split improves over zero-shot but does not surpass Vanilla SICL and trails MetaSICL; worse, despite both being children's speech, it degrades the MyST corpus---a direct symptom of distribution mismatch. Fine-tuning instead on high-resource Common Voice English helps children's ASR in both settings yet still trails MetaSICL, underscoring that explicitly training for in-context behavior beats supervised fine-tuning alone. Overall, narrowly fitting scarce or domain-shifted data over-specializes and hurts generalization, whereas using limited in-domain data as demonstrations adapts more robustly.


\begin{table}[!ht]
\centering
\ninept
\renewcommand{\arraystretch}{1.12}
\setlength{\tabcolsep}{3pt}
\caption{Recipe comparison on the two anchor languages (FLEURS, CER\,$\downarrow$, omnilingual-style normalization). Each method is evaluated at the inference setting matching its training format (0-shot, or few-shot with $k{=}3$ demonstrations). Best per column in \textbf{bold}.}
\label{tab:lowres-ablation}
\begin{tabularx}{\linewidth}{@{}>{\raggedright\arraybackslash}p{6.0cm} c Y Y@{}}
\toprule
& \shortstack{Few-\\shot}
& \multicolumn{2}{c}{$\downarrow$CER} \\
\cmidrule(l{2pt}r{2pt}){3-4}
Method &  & Swahili & Nepali \\
\midrule
\rowcolor{white}
Raw Qwen2.5-Omni                                   & \xmark & 52.9 & 70.7 \\
\rowcolor{gray!15}
MetaSICL (out-of-domain only)                      & \cmark & 51.9 & 70.3 \\
\cmidrule[\heavyrulewidth]{1-4}
\rowcolor{white}
in-domain SFT (0-shot)                             & \xmark & 39.9 & 65.4 \\
\rowcolor{gray!15}
in-domain SFT (SICL)                               & \cmark & 35.6 & 64.0 \\
\rowcolor{white}
in-domain GRPO (no MetaSICL)                       & \xmark & 27.6 & 59.9 \\
\rowcolor{gray!15}
in-domain SICL GRPO (no MetaSICL)                  & \cmark & 92.9 & 60.4 \\
\rowcolor{white}
in-domain SICL GRPO \textbf{(ours)} & \cmark & \textbf{25.6} & \textbf{58.5} \\
\bottomrule
\end{tabularx}
\end{table}

\section{Globalizing ASR to Underserved Low-Resource Languages}
\label{sec:lowres-grpo}

Low-resource automatic speech recognition (ASR) is one of the most consequential tasks for globalizing AI: most of the world's languages are under-served by systems built on high-resource, English-centric data. We therefore ask two questions of MetaSICL in this regime: (1) does the in-context adaptation it strengthens \emph{generalize} to low-resource ASR; and (2) when a small amount of in-domain data \emph{can} be collected---far too little for conventional fine-tuning, but not nothing---how can we best use it to improve further? To answer these, we first explore different ways of combining out-of-domain MetaSICL with in-domain training on two anchor languages, identify the best recipe, and then apply that recipe to five typologically diverse languages.

\subsection{Setup}
We work within FLEURS \citep{conneau2023fleurs}: for each language we use only its dev split as queries ($\sim$200--430 utterances) with SONAR-retrieved in-context examples drawn from the train split, and report character error rate (CER) on the test split (omnilingual-style normalization). We first compare adaptation strategies on two anchor languages---Swahili (\texttt{sw\_ke}; Bantu) and Nepali (\texttt{ne\_np}; Indo-Aryan)---and then validate the best recipe on three further languages: Telugu (\texttt{te\_in}) and Tamil (\texttt{ta\_in}; Dravidian), and Gujarati (\texttt{gu\_in}; Indo-Aryan). Qwen2.5-Omni is initially weak on all five. The strategies we compare span: out-of-domain adaptation only (MetaSICL); \emph{direct} in-domain SFT (the brittle baseline of the preceding case study); and in-domain SICL GRPO \citep{shao2024deepseekmath}, with and without the MetaSICL warmup.

\subsection{Exploring how to best use a small in-domain set}
\label{sec:lowres-ablation}

On the two anchor languages we compare ways of spending the in-domain data against the out-of-domain-only baselines (Table~\ref{tab:lowres-ablation}); each method is reported at the inference setting matching its design/training format (0-shot, or few-shot with $k{=}3$ demonstrations). Three conclusions hold. \textbf{(i)~In-domain data helps far beyond out-of-domain adaptation.} MetaSICL trained only on out-of-domain data reaches 51.9 on Swahili; adding in-domain GRPO on top drives it to 25.6 (Nepali 70.3$\rightarrow$58.5). \textbf{(ii)~In-domain GRPO beats in-domain SFT.} Given the \emph{same} small in-domain set, in-domain GRPO (25.6 on Swahili) far surpasses direct SFT in either 0-shot (39.9) or SICL (35.6) format. \textbf{(iii)~The MetaSICL warmup is what makes in-domain RL work.} Running the identical SICL GRPO \emph{directly from the raw base}---without the MetaSICL warmup---destabilizes and collapses on Swahili (92.9 CER, worse than no adaptation at all), whereas the MetaSICL-initialized policy trains stably to 25.6. A 0-shot GRPO from the raw base is stable (27.6) and a strong simple variant, but the MetaSICL\,$\rightarrow$\,few-shot recipe is best overall. We therefore adopt \textbf{MetaSICL\,$\rightarrow$\,in-domain SICL GRPO} (few-shot) as our recipe and carry it forward.

\begin{table}[!ht]
\centering
\ninept
\renewcommand{\arraystretch}{1.12}
\setlength{\tabcolsep}{3pt}
\caption{Adapting Qwen2.5-Omni to five low-resource FLEURS languages with a small in-domain set (CER\,$\downarrow$, omnilingual-style normalization), comparing the raw model, direct SFT, and our recipe at 0- and 3-shot inference. Best result per language in \textbf{bold}.}
\label{tab:lowres}
\begin{tabularx}{\linewidth}{@{}>{\raggedright\arraybackslash}p{2.7cm} *{6}{Y}@{}}
\toprule
& \multicolumn{2}{c}{raw} & \multicolumn{2}{c}{direct SFT} & \multicolumn{2}{c}{\textbf{ours}} \\
\cmidrule(l{2pt}r{2pt}){2-3}\cmidrule(l{2pt}r{2pt}){4-5}\cmidrule(l{2pt}r{2pt}){6-7}
Language & 0-shot & 3-shot & 0-shot & 3-shot & 0-shot & 3-shot \\
\midrule
\rowcolor{white}
Swahili (\texttt{sw\_ke})   & 52.9 & 47.7 & 39.9 & 40.0 & 29.1 & \textbf{25.6} \\
\rowcolor{gray!15}
Nepali (\texttt{ne\_np})    & 70.7 & 69.5 & 65.4 & 65.6 & 60.4 & \textbf{58.5} \\
\rowcolor{white}
Telugu (\texttt{te\_in})    & 81.0 & 78.9 & 77.1 & 76.7 & 75.5 & \textbf{74.5} \\
\rowcolor{gray!15}
Tamil (\texttt{ta\_in})     & 80.5 & 78.1 & 74.7 & 74.8 & 72.8 & \textbf{71.9} \\
\rowcolor{white}
Gujarati (\texttt{gu\_in})  & 81.8 & 76.1 & 74.4 & 74.5 & 73.4 & \textbf{71.3} \\
\midrule
\rowcolor{gray!15}
Average                     & 73.4 & 70.1 & 66.3 & 66.3 & 62.2 & \textbf{60.4} \\
\bottomrule
\end{tabularx}
\end{table}

\subsection{Generalizing the best recipe to five languages}
Applying our recipe to all five languages (Table~\ref{tab:lowres}), it attains the best CER on \emph{every} language and on average under 3-shot inference, consistently surpassing both the raw model and direct fine-tuning (e.g., Swahili $47.7\!\rightarrow\!25.6$; average $70.1\!\rightarrow\!60.4$). These decent, consistent gains hold across five scripts and three language families, indicating the recipe is not tuned to a particular orthography or family but offers a general, data-efficient route to extending an auditory LLM to languages it was never built to serve.

\section{Conclusion}
This paper studies low-resource adaptation as a core problem for globalizing auditory generative AI. We first show that speech in-context learning (SICL) provides a simple inference-time adaptation mechanism: conditioning on a few audio demonstrations improves children's ASR, multilingual ASR, speech translation, and audio understanding/reasoning without parameter updates. We then propose \textbf{MetaSICL}, a meta-training recipe that uses high-resource speech tasks to train auditory LLMs to use such demonstrations more effectively. Across two model families, MetaSICL strengthens SICL behavior and transfers to low-resource speech and audio settings outside the post-training mixture. Our ablations further show that training-task composition matters, with ASR$+$ST offering the best overall trade-off. We further examine low-resource language ASR as a case study, since speech recognition for underserved languages is central to globalizing generative AI. When a small in-domain corpus is available, MetaSICL provides an effective warmup for data-efficient in-domain reinforcement learning, outperforming direct fine-tuning across five typologically diverse languages. These results suggest that high-resource data can be used to train a general demonstration-conditioned adaptation capability, reducing reliance on large target-domain corpora for every low-resource community.

\section*{Limitations}
Our study covers two model families and a fixed set of benchmarks and retrieval choices; we do not characterize inference-cost scaling with longer contexts or provide extensive qualitative failure analysis. SICL also depends on retrieval quality and the availability of representative demonstrations, which may be scarce in truly data-poor deployments.

\section*{Acknowledgments}
This material is based upon work supported under the AI Research Institutes program by National Science Foundation and the Institute of Education Sciences, U.S. Department of Education, through Award \#2229873 - National AI Institute for Exceptional Education. Any opinions, findings and conclusions or recommendations expressed in this material are those of the author(s) and do not necessarily reflect the views of the National Science Foundation, the Institute of Education Sciences, or the U.S. Department of Education.

This work used the Delta system at the National Center for Supercomputing Applications through allocation beiq-delta-gpu from the Advanced Cyberinfrastructure Coordination Ecosystem: Services \& Support (ACCESS) program, which is supported by National Science Foundation grants \#2138259, \#2138286, \#2138307, \#2137603, and \#2138296.

\section*{Ethical Statement}
We study \emph{speech in-context learning} and post-training for large speech models, with experiments on ASR (including child speech), multilingual ASR, speech translation, and audio understanding/reasoning.

\textbf{Data use and privacy.}
All experiments use existing datasets under their original licenses; some include child speech. We collect no new human-subjects data, rely on the providers' consent and de-identification, attempt no speaker identification or re-identification, and report only aggregate metrics. Any code release will exclude audio or metadata that could re-identify participants.

\textbf{Potential risks and mitigations.}
Better speech recognition aids accessibility and education but can also enable surveillance or profiling, and ASR errors fall disproportionately on child, accented, and low-resource speech. We therefore evaluate on diverse settings to surface such gaps, caution that our results do not imply readiness for safety-critical or rights-impacting use, and recommend informed consent, security controls, and monitoring of differential error rates in any deployment.

\textbf{Misuse considerations.}
Low-data adaptation could lower the barrier to misuse; we focus on research settings, document limitations under domain shift, and encourage responsible release. The models provide no clinical or diagnostic judgments and must not replace professional assessment.

\textbf{Environmental impact.}
We limit compute via parameter-efficient adaptation (LoRA) and bounded-scale experiments, and report key training configurations to aid reproducibility and compute estimation.

\textbf{AI usage statement.}
Generative AI tools assisted only with writing and editing (clarity, grammar, \LaTeX{}). All technical content---design, implementation, results, and claims---was produced and verified by the authors, who take full responsibility; the tools were not used to generate or alter data, labels, or metrics.

\textbf{Software and packages.}
  Because the inference stack can shift results for the same checkpoint, our Qwen2.5-Omni numbers should be read as tied to one setup: \texttt{ms-swift} 4.0.4 with vLLM 0.19.1, \texttt{transformers} 4.57.6, PyTorch 2.10.0 (CUDA 12.8), and Python 3.11.15, one GPU per job on NVIDIA A40/A100.

\textbf{Descriptive statistics.}
We report corpus-level WER/BLEU/accuracy on the full evaluation sets; unless noted, each number is a single run of a fixed checkpoint with fixed decoding/retrieval (no multi-seed mean/std).


\bibliographystyle{colm2026_conference}
\bibliography{custom}

@inproceedings{min2022metaicl,
  title={Metaicl: Learning to learn in context},
  author={Min, Sewon and Lewis, Mike and Zettlemoyer, Luke and Hajishirzi, Hannaneh},
  booktitle={Proceedings of the 2022 conference of the North American chapter of the Association for Computational Linguistics: Human Language Technologies},
  pages={2791--2809},
  year={2022}
}

@misc{zheng2025ticltextembeddingknnspeech,
      title={{TICL: Text-Embedding KNN For Speech In-Context Learning Unlocks Speech Recognition Abilities of Large Multimodal Models}}, 
      author={Haolong Zheng and Yekaterina Yegorova and Mark Hasegawa-Johnson},
      year={2025},
      eprint={2509.13395},
      archivePrefix={arXiv},
      primaryClass={eess.AS},
      url={https://arxiv.org/abs/2509.13395}, 
}

@inproceedings{zhou2025m2r,
  title={{M2R-Whisper: Multi-stage and Multi-scale Retrieval Augmentation for Enhancing Whisper}},
  author={Zhou, Jiaming and Zhao, Shiwan and He, Jiabei and Wang, Hui and Zeng, Wenjia and Chen, Yong and Sun, Haoqin and Kong, Aobo and Qin, Yong},
  booktitle={ICASSP},
  pages={1--5},
  year={2025},
  organization={IEEE}
}

@inproceedings{wang2024can,
  title={{Can Whisper Perform Speech-Based In-Context Learning?}},
  author={Wang, Siyin and Yang, Chao-Han and Wu, Ji and Zhang, Chao},
  booktitle={ICASSP},
  pages={13421--13425},
  year={2024},
  organization={IEEE}
}

@misc{zheng2025ticlcasestudyspeech,
      title={{TICL+: A Case Study On Speech In-Context Learning for Children's Speech Recognition}}, 
      author={Haolong Zheng and Yekaterina Yegorova and Mark Hasegawa-Johnson},
      year={2025},
      eprint={2512.18263},
      archivePrefix={arXiv},
      primaryClass={eess.AS},
      url={https://arxiv.org/abs/2512.18263}, 
}

@misc{coreteam2025mimoaudio,
      title={{MiMo-Audio: Audio Language Models are Few-Shot Learners}}, 
      author={LLM-Core-Team Xiaomi},
      year={2025},
      url={https://github.com/XiaomiMiMo/MiMo-Audio}, 
}

@inproceedings{brown2020language,
 author = {Brown, Tom and Mann, Benjamin and Ryder, Nick and Subbiah, Melanie and Kaplan, Jared D and Dhariwal, Prafulla and Neelakantan, Arvind and Shyam, Pranav and Sastry, Girish and Askell, Amanda and others},
 booktitle = {Advances in Neural Information Processing Systems},
 pages = {1877--1901},
 title = {{Language Models are Few-Shot Learners}},
 volume = {33},
 year = {2020}
}

@inproceedings{huang2023language,
  title={{Language Is Not All You Need: Aligning Perception with Language Models}},
  author={Huang, Shaohan and Dong, Li and Wang, Wenhui and Hao, Yaru and Singhal, Saksham and Ma, Shuming and Lv, Tengchao and Cui, Lei and Mohammed, Owais Khan and Patra, Barun and others},
  booktitle={Advances in Neural Information Processing Systems},
  volume={36},
  pages={72096--72109},
  year={2023}
}

@inproceedings{conneau2023fleurs,
  title={{FLEURS: Few-shot Learning Evaluation of Universal Representations of Speech}},
  author={Conneau, Alexis and Ma, Min and Khanuja, Simran and Zhang, Yu and Axelrod, Vera and Dalmia, Siddharth and Riesa, Jason and Rivera, Clara and Bapna, Ankur},
  booktitle={2022 IEEE Spoken Language Technology Workshop (SLT)},
  pages={798--805},
  year={2023},
  organization={IEEE}
}

@article{shao2024deepseekmath,
  title={{DeepSeekMath: Pushing the Limits of Mathematical Reasoning in Open Language Models}},
  author={Shao, Zhihong and Wang, Peiyi and Zhu, Qihao and Xu, Runxin and Song, Junxiao and Bi, Xiao and Zhang, Haowei and Zhang, Mingchuan and Li, YK and Wu, Yu and Guo, Daya},
  journal={arXiv preprint arXiv:2402.03300},
  year={2024}
}

@inproceedings{chen2024salm,
  title={{SALM: Speech-augmented Language Model with In-context Learning for Speech Recognition and Translation}},
  author={Chen, Zhehuai and Huang, He and Andrusenko, Andrei and Hrinchuk, Oleksii and Puvvada, Krishna C and Li, Jason and Ghosh, Subhankar and Balam, Jagadeesh and Ginsburg, Boris},
  booktitle={ICASSP},
  pages={13521--13525},
  year={2024},
  organization={IEEE}
}

@inproceedings{kong2024audio,
  title={{Audio Flamingo: A Novel Audio Language Model with Few-Shot Learning and Dialogue Abilities}},
  author={Kong, Zhifeng and Goel, Arushi and Badlani, Rohan and Ping, Wei and Valle, Rafael and Catanzaro, Bryan},
  booktitle={Proceedings of the 41st International Conference on Machine Learning},
  pages={25125--25148},
  year={2024}
}

@inproceedings{ihori2025few,
  author={Ihori, Mana and Yamane, Taiga and Kawata, Naotaka and Makishima, Naoki and Tanaka, Tomohiro and Suzuki, Satoshi and Orihashi, Shota and Masumura, Ryo},
  booktitle={ASRU}, 
  title={{Few-shot Personalization via In-Context Learning for Speech Emotion Recognition based on Speech-Language Model}}, 
  year={2025},
  volume={},
  number={},
  pages={1-6},
}

@article{hsu2024smile,
  title={SMILE: speech meta in-context learning for low-resource language automatic speech recognition},
  author={Hsu, Ming-Hao and Lee, Hung-yi},
  journal={arXiv preprint arXiv:2409.10429},
  year={2024}
}

@inproceedings{wang2020covost,
  title     = {CoVoST 2 and Massively Multilingual Speech Translation},
  author    = {Changhan Wang and Anne Wu and Jiatao Gu and Juan Pino},
  year      = {2021},
  booktitle = {INTERSPEECH},
  pages     = {2247--2251},
  doi       = {10.21437/Interspeech.2021-2027},
  issn      = {2958-1796},
}

@inproceedings{commonvoice:2020,
  author = {Ardila, R. and Branson, M. and Davis, K. and Henretty, M. and Kohler, M. and Meyer, J. and Morais, R. and Saunders, L. and Tyers, F. M. and Weber, G.},
  title = {{Common Voice: A Massively-Multilingual Speech Corpus}},
  booktitle = {Proceedings of the 12th Conference on Language Resources and Evaluation},
  pages = {4211--4215},
  year = {2020}
}

@article{wang2025mmsu,
  title={{MMSU: A Massive Multi-task Spoken Language Understanding and Reasoning Benchmark}},
  author={Wang, Dingdong and Wu, Jincenzi and Li, Junan and Yang, Dongchao and Chen, Xueyuan and Zhang, Tianhua and Meng, Helen},
  journal={arXiv preprint arXiv:2506.04779},
  year={2025}
}

@article{Qwen2.5-Omni,
  title={{Qwen2.5-Omni Technical Report}},
  author={Jin, Xu and Zhifang, Guo and Jinzheng, He and Hangrui, Hu and Ting, He and Shuai, Bai and Keqin, Chen and Jialin, Wang and Yang, Fan and Kai, Dang and Bin, Zhang and Xiong, Wang and Yunfei, Chu and Junyang, Lin},
  journal={arXiv preprint arXiv:2503.20215},
  year={2025}
}

@inproceedings{zheng2025interspeech,
  title     = {{The Interspeech 2025 Speech Accessibility Project Challenge}},
  author    = {Xiuwen Zheng and Bornali Phukon and Jonghwan Na and Ed Cutrell and Kyu J. Han and Mark Hasegawa-Johnson and Pan-Pan Jiang and Aadhrik Kuila and Colin Lea and Bob MacDonald and Gautam Mantena and Venkatesh Ravichandran and Leda Sari and Katrin Tomanek and Chang D. Yoo and Chris Zwilling},
  year      = {2025},
  booktitle = {{INTERSPEECH}},
  pages     = {3269--3273},
  doi       = {10.21437/Interspeech.2025-566},
  issn      = {2958-1796},
}

@inproceedings{pradhan2024my,
  title={{My Science Tutor (MyST) -- A Large Corpus of Children's Conversational Speech}},
  author={Pradhan, Sameer and Cole, Ronald and Ward, Wayne},
  booktitle={Proceedings of the 2024 Joint International Conference on Computational Linguistics, Language Resources and Evaluation},
  pages={12040--12045},
  year={2024}
}

@article{ai4exceptionaled_rsr_hf,
  title={{Diagnostic Accuracy of Sentence Recall and Past Tense Measures for Identifying Children's Language Impairments}},
  author={Redmond, Sean M and Ash, Andrea C and Christopulos, Tyler T and Pfaff, Theresa},
  journal={Journal of Speech, Language, and Hearing Research},
  volume={62},
  number={7},
  pages={2438--2454},
  year={2019},
  publisher={American Speech-Language-Hearing Association}
}

@article{yang2024uniaudio,
  title={Uniaudio 1.5: Large language model-driven audio codec is a few-shot audio task learner},
  author={Yang, Dongchao and Guo, Haohan and Wang, Yuanyuan and Huang, Rongjie and Li, Xiang and Tan, Xu and Wu, Xixin and Meng, Helen},
  journal={NeurIPS},
  pages={56802--56827},
  year={2024}
}

@article{pan2023cosmic,
  title={Cosmic: Data efficient instruction-tuning for speech in-context learning},
  author={Pan, Jing and Wu, Jian and Gaur, Yashesh and Sivasankaran, Sunit and Chen, Zhuo and Liu, Shujie and Li, Jinyu},
  journal={arXiv preprint arXiv:2311.02248},
  year={2023}
}

@inproceedings{wang2024bayesian,
  title={Bayesian Example Selection Improves In-Context Learning for Speech, Text and Visual Modalities},
  author={Wang, Siyin and Yang, Chao-Han and Wu, Ji and Zhang, Chao},
  booktitle={EMNLP},
  pages={20812--20828},
  year={2024}
}

\newpage
\appendix

\section{MetaSICL Algorithm}

\begin{algorithm}[ht]
\caption{MetaSICL}
\label{alg:training-overview}
\DontPrintSemicolon

\KwIn{$C$ training tasks, \\ each has query set $\mathcal{D}_{query}^{(c)}=\{(x_c,y_c)^{(j)}\}_{j=1}^{N_{c,query}}$, and demo pool $\mathcal{D}_{pool}^{(c)}=\{(x_c,y_c)^{(j)}\}_{j=1}^{N_{c,pool}}$}

\For{$step \gets 1$ \KwTo $\textit{Total\_Step}$}{
  \textbf{(1)} Sample task index $c \sim \{1,\ldots,C\}$\;
  \textbf{(2)} Sample $(x_{\text{query}},y_{\text{query}})\sim \mathcal{D}^{(c)}_{\text{query}}$\;
  \textbf{(3)} Retrieve $\{(x_i,y_i)\}_{i=1}^{k}$ from $\mathcal{D}^{(c)}_{\text{pool}}$\;
  \textbf{(4)} Update parameters to maximize
  $P\!\left(y_{\text{query}} \mid x_1,y_1,\ldots,x_k,y_k,x_{\text{query}}\right)$\;
}
\end{algorithm}

\section{General Audio Understanding and Reasoning Performance Breakdowns}
\label{appendix:AU-AR}


\begin{table*}[tbp]
\centering
\small
\renewcommand{\arraystretch}{1.12}
\setlength{\tabcolsep}{5pt}
\caption{MMAU accuracy breakdown by group/item for \textbf{MiMo-Audio}.}
\label{tab:mmau_breakdown_mimo}

\resizebox{\linewidth}{!}{%
\begin{tabular}{l p{6.8cm} c ccccc}
\toprule
\textbf{Group} & \textbf{Item} & \textbf{n} &
\textbf{0shot} & \textbf{ICL} & \textbf{ASR} & \textbf{ASR$+$ST} & \textbf{$+$SQA} \\
\midrule
Task & sound & 333 & 71.77\% & 75.98\% & 78.98\% & 75.98\% & 78.08\% \\
Task & music & 334 & 65.27\% & 66.77\% & 65.57\% & 68.86\% & 68.26\% \\
Task & speech & 333 & 63.66\% & 75.08\% & 71.17\% & 73.87\% & 73.87\% \\
\midrule
Difficulty & easy & 224 & 55.80\% & 66.52\% & 59.82\% & 64.29\% & 65.18\% \\
Difficulty & hard & 236 & 64.41\% & 71.19\% & 72.88\% & 71.61\% & 69.92\% \\
Difficulty & medium & 540 & 72.59\% & 75.74\% & 76.48\% & 77.04\% & 78.33\% \\
\midrule
Sub-category & Acoustic Source Inference & 48 & 72.92\% & 81.25\% & 85.42\% & 70.83\% & 79.17\% \\
Sub-category & Temporal Event Reasoning & 48 & 66.67\% & 62.50\% & 75.00\% & 62.50\% & 66.67\% \\
Sub-category & Dissonant Emotion Interpretation & 35 & 80.00\% & 85.71\% & 74.29\% & 82.86\% & 80.00\% \\
Sub-category & Event-Based Knowledge Retrieval & 33 & 75.76\% & 90.91\% & 81.82\% & 93.94\% & 87.88\% \\
Sub-category & Counting & 29 & 48.28\% & 55.17\% & 55.17\% & 62.07\% & 55.17\% \\
Sub-category & Phonemic Stress Pattern Analysis & 53 & 39.62\% & 62.26\% & 52.83\% & 50.94\% & 60.38\% \\
Sub-category & Emotion State summarisation & 44 & 56.82\% & 56.82\% & 61.36\% & 61.36\% & 63.64\% \\
Sub-category & Conversational Fact Retrieval & 22 & 86.36\% & 95.45\% & 95.45\% & 90.91\% & 100.00\% \\
Sub-category & Key highlight Extraction & 21 & 80.95\% & 90.48\% & 95.24\% & 90.48\% & 90.48\% \\
Sub-category & Multi Speaker Role Mapping & 27 & 100.00\% & 100.00\% & 100.00\% & 100.00\% & 100.00\% \\
Sub-category & Phonological Sequence Decoding & 49 & 59.18\% & 81.63\% & 69.39\% & 75.51\% & 71.43\% \\
Sub-category & Emotion Flip Detection & 20 & 35.00\% & 45.00\% & 55.00\% & 55.00\% & 50.00\% \\
Sub-category & Instrumentation & 35 & 60.00\% & 71.43\% & 68.57\% & 77.14\% & 68.57\% \\
Sub-category & Temporal Reasoning & 56 & 41.07\% & 37.50\% & 39.29\% & 41.07\% & 39.29\% \\
Sub-category & Lyrical Reasoning & 10 & 90.00\% & 90.00\% & 90.00\% & 90.00\% & 90.00\% \\
Sub-category & Socio-cultural Interpretation & 20 & 65.00\% & 75.00\% & 70.00\% & 70.00\% & 80.00\% \\
Sub-category & Rhythm and Tempo Understanding & 46 & 69.57\% & 63.04\% & 60.87\% & 65.22\% & 71.74\% \\
Sub-category & Musical Texture Interpretation & 34 & 73.53\% & 76.47\% & 70.59\% & 76.47\% & 76.47\% \\
Sub-category & Melodic Structure Interpretation & 33 & 66.67\% & 66.67\% & 54.55\% & 63.64\% & 54.55\% \\
Sub-category & Harmony and Chord Progressions & 33 & 63.64\% & 63.64\% & 66.67\% & 69.70\% & 63.64\% \\
Sub-category & Musical Genre Reasoning & 34 & 70.59\% & 79.41\% & 88.24\% & 85.29\% & 88.24\% \\
Sub-category & Event-Based Sound Reasoning & 48 & 79.17\% & 83.33\% & 81.25\% & 89.58\% & 81.25\% \\
Sub-category & Emotional Tone Interpretation & 33 & 84.85\% & 84.85\% & 84.85\% & 84.85\% & 87.88\% \\
Sub-category & Eco-Acoustic Knowledge & 47 & 68.09\% & 72.34\% & 76.60\% & 82.98\% & 78.72\% \\
Sub-category & Ambient Sound Interpretation & 48 & 62.50\% & 72.92\% & 77.08\% & 72.92\% & 75.00\% \\
Sub-category & Acoustic Scene Reasoning & 48 & 70.83\% & 70.83\% & 64.58\% & 64.58\% & 75.00\% \\
Sub-category & Sound-Based Event Recognition & 46 & 82.61\% & 89.13\% & 93.48\% & 89.13\% & 91.30\% \\
\midrule
Overall & Total Accuracy & 1000 & 66.90\% & 72.60\% & 71.90\% & 72.90\% & 73.40\% \\
\bottomrule
\end{tabular}%
}
\end{table*}

\begin{table*}[tbp]
\centering
\small
\renewcommand{\arraystretch}{1.12}
\setlength{\tabcolsep}{5pt}
\caption{MMAU accuracy breakdown by group/item for \textbf{Qwen2.5-Omni}.}
\label{tab:mmau_breakdown_qwen}

\resizebox{\linewidth}{!}{%
\begin{tabular}{l p{6.8cm} c ccccc}
\toprule
\textbf{Group} & \textbf{Item} & \textbf{n} &
\textbf{0shot} & \textbf{ICL} & \textbf{ASR} & \textbf{ASR$+$ST} & \textbf{$+$SQA} \\
\midrule
Task & sound & 333 & 68.17\% & 73.87\% & 77.18\% & 75.08\% & 78.08\% \\
Task & music & 334 & 62.28\% & 62.28\% & 61.98\% & 65.27\% & 66.77\% \\
Task & speech & 333 & 66.97\% & 65.77\% & 69.67\% & 72.97\% & 71.47\% \\
\midrule
Difficulty & easy & 224 & 63.84\% & 61.16\% & 63.39\% & 68.30\% & 68.30\% \\
Difficulty & hard & 236 & 61.44\% & 60.17\% & 64.83\% & 63.56\% & 69.07\% \\
Difficulty & medium & 540 & 68.52\% & 72.96\% & 74.26\% & 75.56\% & 75.00\% \\
\midrule
Sub-category & Acoustic Source Inference & 48 & 81.25\% & 81.25\% & 93.75\% & 89.58\% & 87.50\% \\
Sub-category & Temporal Event Reasoning & 48 & 41.67\% & 64.58\% & 52.08\% & 58.33\% & 60.42\% \\
Sub-category & Dissonant Emotion Interpretation & 35 & 71.43\% & 88.57\% & 82.86\% & 94.29\% & 82.86\% \\
Sub-category & Event-Based Knowledge Retrieval & 33 & 75.76\% & 75.76\% & 78.79\% & 87.88\% & 78.79\% \\
Sub-category & Counting & 29 & 68.97\% & 58.62\% & 58.62\% & 62.07\% & 55.17\% \\
Sub-category & Phonemic Stress Pattern Analysis & 53 & 43.40\% & 41.51\% & 43.40\% & 49.06\% & 47.17\% \\
Sub-category & Emotion State summarisation & 44 & 50.00\% & 45.45\% & 56.82\% & 45.45\% & 54.55\% \\
Sub-category & Conversational Fact Retrieval & 22 & 95.45\% & 86.36\% & 90.91\% & 95.45\% & 90.91\% \\
Sub-category & Key highlight Extraction & 21 & 76.19\% & 85.71\% & 80.95\% & 85.71\% & 85.71\% \\
Sub-category & Multi Speaker Role Mapping & 27 & 100.00\% & 100.00\% & 100.00\% & 100.00\% & 100.00\% \\
Sub-category & Phonological Sequence Decoding & 49 & 79.59\% & 77.55\% & 89.80\% & 91.84\% & 89.80\% \\
Sub-category & Emotion Flip Detection & 20 & 25.00\% & 10.00\% & 20.00\% & 30.00\% & 45.00\% \\
Sub-category & Instrumentation & 35 & 74.29\% & 62.86\% & 74.29\% & 77.14\% & 77.14\% \\
Sub-category & Temporal Reasoning & 56 & 42.86\% & 35.71\% & 35.71\% & 41.07\% & 37.50\% \\
Sub-category & Lyrical Reasoning & 10 & 60.00\% & 90.00\% & 90.00\% & 90.00\% & 100.00\% \\
Sub-category & Socio-cultural Interpretation & 20 & 70.00\% & 65.00\% & 70.00\% & 65.00\% & 70.00\% \\
Sub-category & Rhythm and Tempo Understanding & 46 & 50.00\% & 54.35\% & 56.52\% & 58.70\% & 45.65\% \\
Sub-category & Musical Texture Interpretation & 34 & 70.59\% & 76.47\% & 64.71\% & 73.53\% & 67.65\% \\
Sub-category & Melodic Structure Interpretation & 33 & 57.58\% & 63.64\% & 57.58\% & 57.58\% & 72.73\% \\
Sub-category & Harmony and Chord Progressions & 33 & 51.52\% & 60.61\% & 63.64\% & 63.64\% & 69.70\% \\
Sub-category & Musical Genre Reasoning & 34 & 88.24\% & 82.35\% & 70.59\% & 79.41\% & 88.24\% \\
Sub-category & Event-Based Sound Reasoning & 48 & 66.67\% & 77.08\% & 83.33\% & 79.17\% & 81.25\% \\
Sub-category & Emotional Tone Interpretation & 33 & 75.76\% & 72.73\% & 78.79\% & 81.82\% & 90.91\% \\
Sub-category & Eco-Acoustic Knowledge & 47 & 74.47\% & 78.72\% & 80.85\% & 85.11\% & 82.98\% \\
Sub-category & Ambient Sound Interpretation & 48 & 77.08\% & 70.83\% & 81.25\% & 68.75\% & 79.17\% \\
Sub-category & Acoustic Scene Reasoning & 48 & 58.33\% & 62.50\% & 66.67\% & 64.58\% & 75.00\% \\
Sub-category & Sound-Based Event Recognition & 46 & 78.26\% & 82.61\% & 82.61\% & 80.43\% & 80.43\% \\
\midrule
Overall & Total Accuracy & 1000 & 65.80\% & 67.30\% & 69.60\% & 71.10\% & 72.10\% \\
\bottomrule
\end{tabular}%
}
\end{table*}




\begin{table*}[tbp]
\centering
\small
\renewcommand{\arraystretch}{1.12}
\setlength{\tabcolsep}{5pt}
\caption{MMAR Accuracy breakdown for \textbf{MiMo-Audio}.}
\label{tab:mmar_breakdown_mimo}

\resizebox{\linewidth}{!}{%
\begin{tabular}{l p{6.4cm} c ccccc}
\toprule
\textbf{Group} & \textbf{Item} & \textbf{n} &
\textbf{0shot} & \textbf{Vanilla SICL} & \textbf{ASR} & \textbf{ASR$+$ST} & \textbf{$+$SQA} \\
\midrule
Modality & sound & 165 & 53.33\% & 52.73\% & 55.15\% & 60.61\% & 62.42\% \\
Modality & music & 206 & 39.32\% & 43.20\% & 44.17\% & 46.60\% & 46.12\% \\
Modality & speech & 294 & 58.84\% & 63.61\% & 63.61\% & 66.33\% & 66.67\% \\
Modality & mix-sound-music & 11 & 45.45\% & 27.27\% & 27.27\% & 27.27\% & 45.45\% \\
Modality & mix-sound-speech & 218 & 63.30\% & 68.35\% & 65.60\% & 69.27\% & 67.43\% \\
Modality & mix-music-speech & 82 & 58.54\% & 57.32\% & 56.10\% & 58.54\% & 60.98\% \\
Modality & mix-sound-music-speech & 24 & 58.33\% & 83.33\% & 66.67\% & 70.83\% & 75.00\% \\
\midrule
Category & Signal Layer & 43 & 53.49\% & 55.81\% & 48.84\% & 62.79\% & 60.47\% \\
Category & Perception Layer & 404 & 52.72\% & 53.71\% & 56.19\% & 57.92\% & 58.42\% \\
Category & Semantic Layer & 412 & 58.98\% & 65.05\% & 62.38\% & 66.50\% & 66.26\% \\
Category & Cultural Layer & 141 & 48.23\% & 51.77\% & 51.06\% & 53.19\% & 56.03\% \\
\midrule
Sub-category & Speaker Analysis & 48 & 62.50\% & 62.50\% & 54.17\% & 64.58\% & 60.42\% \\
Sub-category & Environmental Perception and Reasoning & 149 & 59.06\% & 61.07\% & 66.44\% & 71.14\% & 73.15\% \\
Sub-category & Content Analysis & 304 & 60.20\% & 67.76\% & 64.47\% & 67.43\% & 68.42\% \\
Sub-category & Correlation Analysis & 50 & 62.00\% & 58.00\% & 60.00\% & 58.00\% & 62.00\% \\
Sub-category & Counting and Statistics & 99 & 42.42\% & 44.44\% & 40.40\% & 39.39\% & 39.39\% \\
Sub-category & Professional Knowledge and Reasoning & 71 & 47.89\% & 54.93\% & 53.52\% & 52.11\% & 56.34\% \\
Sub-category & Culture of Speaker & 52 & 50.00\% & 46.15\% & 48.08\% & 57.69\% & 59.62\% \\
Sub-category & Aesthetic Evaluation & 8 & 37.50\% & 62.50\% & 37.50\% & 50.00\% & 50.00\% \\
Sub-category & Emotion and Intention & 60 & 50.00\% & 53.33\% & 58.33\% & 63.33\% & 60.00\% \\
Sub-category & Anomaly Detection & 17 & 58.82\% & 52.94\% & 47.06\% & 64.71\% & 64.71\% \\
Sub-category & Spatial Analysis & 15 & 60.00\% & 53.33\% & 73.33\% & 73.33\% & 66.67\% \\
Sub-category & Temporal Analysis & 28 & 53.57\% & 57.14\% & 53.57\% & 57.14\% & 57.14\% \\
Sub-category & Acoustic Quality Analysis & 18 & 33.33\% & 44.44\% & 33.33\% & 44.44\% & 44.44\% \\
Sub-category & Music Theory & 63 & 44.44\% & 46.03\% & 50.79\% & 52.38\% & 49.21\% \\
Sub-category & Audio Difference Analysis & 8 & 87.50\% & 87.50\% & 87.50\% & 100.00\% & 87.50\% \\
Sub-category & Imagination & 10 & 50.00\% & 50.00\% & 60.00\% & 40.00\% & 40.00\% \\
\midrule
Overall & Total Accuracy & 1000 & 54.70\% & 58.20\% & 57.70\% & 61.00\% & 61.40\% \\
\bottomrule
\end{tabular}%
}
\end{table*}

\begin{table*}[tbp]
\centering
\small
\renewcommand{\arraystretch}{1.12}
\setlength{\tabcolsep}{5pt}
\caption{MMAR Accuracy breakdown for \textbf{Qwen2.5-Omni}.}
\label{tab:mmar_breakdown_qwen}

\resizebox{\linewidth}{!}{%
\begin{tabular}{l p{6.4cm} c ccccc}
\toprule
\textbf{Group} & \textbf{Item} & \textbf{n} &
\textbf{0shot} & \textbf{Vanilla SICL} & \textbf{ASR} & \textbf{ASR$+$ST} & \textbf{$+$SQA} \\
\midrule
Modality & sound & 165 & 47.27\% & 52.12\% & 55.76\% & 59.39\% & 60.00\% \\
Modality & music & 206 & 36.89\% & 39.81\% & 42.23\% & 44.66\% & 41.75\% \\
Modality & speech & 294 & 51.70\% & 56.46\% & 56.12\% & 55.44\% & 57.48\% \\
Modality & mix-sound-music & 11 & 27.27\% & 63.64\% & 45.45\% & 36.36\% & 54.55\% \\
Modality & mix-sound-speech & 218 & 56.42\% & 58.72\% & 57.34\% & 55.96\% & 56.42\% \\
Modality & mix-music-speech & 82 & 54.88\% & 63.41\% & 63.41\% & 60.98\% & 58.54\% \\
Modality & mix-sound-music-speech & 24 & 62.50\% & 70.83\% & 66.67\% & 62.50\% & 58.33\% \\
\midrule
Category & Signal Layer & 43 & 25.58\% & 46.51\% & 51.16\% & 51.16\% & 60.47\% \\
Category & Perception Layer & 404 & 46.29\% & 48.76\% & 52.23\% & 52.72\% & 50.50\% \\
Category & Semantic Layer & 412 & 55.10\% & 59.22\% & 57.04\% & 58.74\% & 60.19\% \\
Category & Cultural Layer & 141 & 47.52\% & 54.61\% & 52.48\% & 47.52\% & 47.52\% \\
\midrule
Sub-category & Speaker Analysis & 48 & 56.25\% & 45.83\% & 52.08\% & 60.42\% & 64.58\% \\
Sub-category & Environmental Perception and Reasoning & 149 & 64.43\% & 64.43\% & 63.76\% & 65.77\% & 66.44\% \\
Sub-category & Content Analysis & 304 & 55.26\% & 62.17\% & 57.89\% & 59.87\% & 60.53\% \\
Sub-category & Correlation Analysis & 50 & 46.00\% & 60.00\% & 54.00\% & 56.00\% & 62.00\% \\
Sub-category & Counting and Statistics & 99 & 31.31\% & 32.32\% & 36.36\% & 35.35\% & 30.30\% \\
Sub-category & Professional Knowledge and Reasoning & 71 & 47.89\% & 56.34\% & 50.70\% & 56.34\% & 47.89\% \\
Sub-category & Culture of Speaker & 52 & 44.23\% & 51.92\% & 53.85\% & 36.54\% & 48.08\% \\
Sub-category & Aesthetic Evaluation & 8 & 75.00\% & 50.00\% & 37.50\% & 37.50\% & 50.00\% \\
Sub-category & Emotion and Intention & 60 & 53.33\% & 55.00\% & 56.67\% & 51.67\% & 55.00\% \\
Sub-category & Anomaly Detection & 17 & 35.29\% & 47.06\% & 47.06\% & 47.06\% & 76.47\% \\
Sub-category & Spatial Analysis & 15 & 53.33\% & 40.00\% & 66.67\% & 53.33\% & 53.33\% \\
Sub-category & Temporal Analysis & 28 & 28.57\% & 42.86\% & 50.00\% & 46.43\% & 42.86\% \\
Sub-category & Acoustic Quality Analysis & 18 & 11.11\% & 44.44\% & 50.00\% & 50.00\% & 55.56\% \\
Sub-category & Music Theory & 63 & 33.33\% & 33.33\% & 46.03\% & 49.21\% & 38.10\% \\
Sub-category & Audio Difference Analysis & 8 & 37.50\% & 50.00\% & 62.50\% & 62.50\% & 37.50\% \\
Sub-category & Imagination & 10 & 40.00\% & 60.00\% & 70.00\% & 50.00\% & 40.00\% \\
\midrule
Overall & Total Accuracy & 1000 & 49.20\% & 53.80\% & 54.20\% & 54.40\% & 54.50\% \\
\bottomrule
\end{tabular}%
}
\end{table*}

\label{mmar}

\begin{table}[tbp]
\centering
\footnotesize
\setlength{\tabcolsep}{3pt}
\renewcommand{\arraystretch}{1.10}

\begin{tabularx}{\columnwidth}{@{}l l >{\raggedright\arraybackslash}X r@{}}
\toprule
Component & Dataset & Task / pair & \multicolumn{1}{c}{\#samples} \\
\midrule
ASR & Common Voice & ASR (en) & 16{,}368 \\
$+$ST & CoVoST2 & +ST (total) & 37{,}087 \\
& & \hspace{1.2em} en$\rightarrow$zh & 15{,}427 \\
& & \hspace{1.2em} de$\rightarrow$en & 13{,}500 \\
& & \hspace{1.2em} zh$\rightarrow$en & 4{,}842 \\
& & \hspace{1.2em} pt$\rightarrow$en & 3{,}318 \\
$+$SQA$^*$ & MMSU & +SQA & 5{,}000 \\
\bottomrule
\end{tabularx}
\caption{Training data. MetaSICL uses the ASR$+$ST mixture; the $+$SQA ablation additionally uses MMSU ($^*$).}
\label{tab:training-data}
\end{table}

\end{document}